\documentclass[final]{cvpr}

\usepackage{times}
\usepackage{epsfig}
\usepackage{graphicx}
\usepackage{amsmath}
\usepackage{amssymb}
\usepackage{CJK}
\usepackage{float}
\usepackage{blindtext}
\usepackage{cuted}
\usepackage{capt-of}
\usepackage{booktabs}
\usepackage{bbding}
\usepackage{amssymb}

\usepackage[pagebackref=true,breaklinks=true,colorlinks,bookmarks=false]{hyperref}

\def\cvprPaperID{75} % *** Enter the CVPR Paper ID here
\def\confYear{CVPR 2021}

\begin{document}

\title{HDRUNet: Single Image HDR Reconstruction with 

Denoising and Dequantization}  % **** Enter the paper title here

\author{
 Xiangyu Chen\textsuperscript{\rm 1} \space \space 
 Yihao Liu\textsuperscript{\rm 1,2} \space \space 
 Zhengwen Zhang\textsuperscript{\rm 1} \space \space 
 Yu Qiao\textsuperscript{\rm 1,3}   \space \space 
 Chao Dong\textsuperscript{\rm 1,4 \thanks{Corresponding author.}} \\ 
 \textsuperscript{\rm 1}Key Laboratory of Human-Machine Intelligence-Synergy Systems,\\Shenzhen Institutes of Advanced Technology, Chinese Academy of Sciences\\
 \textsuperscript{\rm 2}University of Chinese Academy of Sciences \space \space 
 \textsuperscript{\rm 3}Shanghai AI Lab, Shanghai, China\\ 
 \textsuperscript{\rm 4}SIAT Branch, Shenzhen Institute of Artificial Intelligence and Robotics for Society\\ 
 {\tt\small\{chxy95, zhengwen.zhang02\}@gmail.com}\space \space {\tt\small liuyihao14@mails.ucas.ac.cn} \space \space {\tt\small\{yu.qiao, chao.dong\}@siat.ac.cn} 
}

\twocolumn[{%
\renewcommand\twocolumn[1][]{#1}%
\maketitle
\vspace{-20pt}
\begin{center}
    \centering
    %\captionsetup{type=figure}
    \includegraphics[width=1\linewidth]{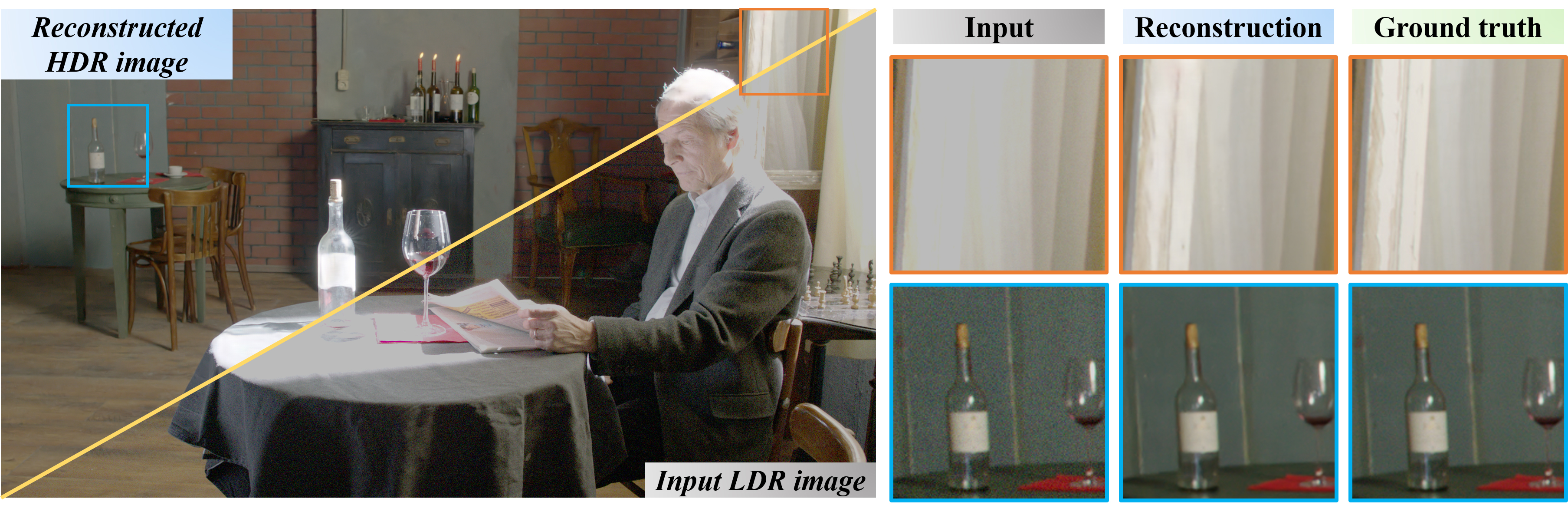}
    \captionof{figure}{HDR reconstruction with denoising and dequantization from a single LDR image. We propose a novel learning based method for single image HDR reconstruction with denoising and dequantization. The proposed method consists of a spatially dynamic encoder-decoder network and a new $Tanh \textunderscore L_1$ loss function. The visual comparison shows that our method reconstructs information in over-exposed regions and also reduces the noise and quantization loss in well-exposed regions. All the images have been $\mu$-law tone-mapped for display. We slightly increase the contrast of patches in the bottom row for clearer visualization. \textbf{Please zoom in for best view.}}
\end{center}%
}]

%%%%%%%%% ABSTRACT
\begin{abstract}

Most consumer-grade digital cameras can only capture a limited range of luminance in real-world scenes due to sensor constraints. Besides, noise and quantization errors are often introduced in the imaging process. In order to obtain high dynamic range (HDR) images with excellent visual quality, the most common solution is to combine multiple images with different exposures. However, it is not always feasible to obtain multiple images of the same scene and most HDR reconstruction methods ignore the noise and quantization loss. In this work, we propose a novel learning-based approach using a spatially dynamic encoder-decoder network, HDRUNet, to learn an end-to-end mapping for single image HDR reconstruction with denoising and dequantization. The network consists of a UNet-style base network to make full use of the hierarchical multi-scale information, a condition network to perform pattern-specific modulation and a weighting network for selectively retaining information. Moreover, we propose a $Tanh \textunderscore L_1$ loss function to balance the impact of over-exposed values and well-exposed values on the network learning. Our method achieves the state-of-the-art performance in quantitative comparisons and visual quality. The proposed HDRUNet model won the second place in the single frame track of NITRE2021 High Dynamic Range Challenge. The code is available at \url{https://github.com/chxy95/HDRUNet}.

\end{abstract}

\vspace{-15pt}

%%%%%%%%% BODY TEXT - ENTER YOUR RESPONSE BELOW
\section{Introduction}

High dynamic range (HDR) images are capable of recording a more realistic appearance of the scene, which can significantly improve the viewing experience. However, limited by the sensor, most consumer-grade digital cameras can only capture a limited range of luminance. In addition, noise and quantization errors are often introduced in the imaging processing. The most commonly used method to generate an HDR image is to merge a set of LDR images captured with different exposures \cite{debevec2008recovering}. However, these approaches have to deal with the object motion among different LDR images \cite{wu2018deep, kalantari2017deep, oh2014robust, hu2013hdr}, and multiple images captured at the same scene are not always feasible. Besides, most HDR reconstruction methods only focus on dynamic range expansion \cite{eilertsen2017hdr, santos2020single} and ignore the noise and quantization loss in the well-exposed regions. 

Single image HDR reconstruction with denoising and dequantization is a challenging problem. First, it is hard to recover the missing details in the under-/over-exposed regions from a single LDR input due to severe information loss. Second, dealing with the problem of joint HDR reconstruction, denoising and dequantization is a challenge for the network design and training. Some traditional single image HDR reconstruction approaches directly improve the brightness or enhance the contrast of the input \cite{masia2009evaluation, masia2017, akyuz2007hdr}. A number of techniques utilize image local heuristics to expand the dynamic range \cite{banterle2006inverse, kovaleski2014high}. Most recent data-driven single image HDR reconstruction methods deal with the problem by recovering the over-exposed regions \cite{eilertsen2017hdr}. Note that these methods are all proposed to predict the linear HDR values in luminance domain and do not explicitly perform denoising. There are also several methods that have been proposed recently, aiming at predicting the non-linear HDR values in display format under the HDR standard \cite{kim2019deep, kim2020jsi}. They also do not consider the denoising issues.  

In this work, we aim to predict a non-linear 16-bit HDR image after gamma correction from a single 8-bit LDR noisy image. We propose a spatially dynamic encoder-decoder network, called HDRUNet, to deal with restoration details in under-/over-exposed regions along with denoising and dequantization for the whole image. We design our approach based on two observations. First, noise and quantization errors certainly exist in LDR images in comparison with their HDR ground truths, and the patterns in over-exposed regions are obviously different from those in well-exposed regions. Second, distributions of noise are spatially variant, which are not uniform like Gaussian white noise. In order to address these issues, we first design a network consisting of three parts, including a UNet-like base network that can utilize multi-scale information, a condition network that performs spatially dynamic modulation for different patterns, and a weighting network for adaptively retaining information of the input. Besides, we propose a new $Tanh \textunderscore L_1$ loss function that normalizes values into [0, 1] to balance the impact of high luminance values and the other values during training, in order to prevent the network from only focusing on high luminance values.

Our contributions are three-fold: 
\begin{itemize}
\setlength{\itemsep}{0pt}
\setlength{\parsep}{0pt}
\setlength{\parskip}{0pt}
\item We propose a new deep network to reconstruct a high quality HDR image with denoising and dequantization from a single LDR image.
\item We introduce a $Tanh \textunderscore L_1$ loss for the task. Compared to the other commonly used losses of image restoration, this loss can lead to better quantitative performance and visual quality.
\item Experiments show that our method outperforms the state-of-the-art methods both quantitatively and qualitatively, and we won the second place in the single frame track of NTIRE2021 HDR Challenge \cite{perez2021ntire}.

\end{itemize}

\section{Related Work}
\subsection{HDR Reconstruction}
The task of image HDR reconstruction, which is also known as inverse tone mapping \cite{banterle2006inverse}, has been extensively studied in the previous decades. The most common technique is to fuse a stack of bracketed exposure LDR images \cite{debevec2008recovering}. There are also recent methods applying CNNs to fuse multiple LDR images \cite{wu2018deep, kalantari2017deep, endo2017deep}. In this paper, we focus on reconstructing HDR image from a single LDR image. 

Traditional single image HDR reconstruction methods exploit internal image characteristics to predict the luminance of the scene. For example, \cite{akyuz2007hdr, banterle2009high, banterle2006inverse, banterle2007framework} estimate the density of light sources to expand the dynamic range and \cite{huo2014physiological, kovaleski2014high} apply cross-bilateral filter to enhance the input LDR images. There are also several approaches \cite{masia2009evaluation, masia2017} using global operator for approximating tone expansion to improve the visual quality. 

Recently CNNs have also shown great performance for image restoration and enhancement tasks such as image super-resolution \cite{dong2014learning}, compression artifact reduction \cite{dong2015compression}, denoising \cite{zhang2017beyond}, photo retouching \cite{he2020conditional} and inpainting \cite{yeh2017semantic}, etc. Several methods have been developed to learn a direct LDR-to-HDR mapping. Eilertsen et al. \cite{eilertsen2017hdr} propose HDRCNN to recover missing details in the over-exposed regions and Santos et al. \cite{santos2020single} improve the method by adding masked features and perceptual loss. However, their methods ignore the quantization artifacts and noise in the well-exposed areas. SingleHDR \cite{liu2020single} learns LDR-to-HDR mapping by reversing the camera pipeline. These approaches aim at predicting the linear HDR luminance. Kim et al. \cite{kim2019deep} propose Deep SR-ITM to solve the problem of joint super-resolution and inverse tone-mapping, while they aim to predict HDR pixel values in display format under HDR standard involving wide color gamut and HDR transfer function. In this work, we focus on the problem of single image HDR reconstruction with denoising and dequantization.

\subsection{Denoising}
Image denoising is a classic topic in the field of low level vision. Traditional methods use various models to model the image prior to achieve denoising, such as \cite{dabov2007image, mairal2009non, dong2012nonlocally, gu2014weighted, weiss2007makes}. These prior-based methods are generally time-consuming and involve manually chosen parameters. Recently, there have been several attempts to preform denoising by CNNs \cite{zhang2017beyond, plotz2018neural, mao2016image}. However, these methods are designed for Gaussian white noise which usually generalize poorly to real-world noisy images \cite{plotz2017benchmarking}. For addressing this issue, several approaches are proposed by taking noise level prior as network input to handle different noise levels and spatially variant noise \cite{zhang2018ffdnet, he2019modulating, he2019interactive}. In this work, we add a spatially dynamic modulation module to perform denoising inside along with HDR reconstruction.

\subsection{Dequantization}
Quantization errors are inevitably occurred in the imaging process. It is reflected in the image as scattered noise and artifacts (e.g. contouring or banding artifacts) in regions with smooth gradient changes. Previous works on bit-depth expansion smooth image by applying the spatially adaptive filter \cite{daly2004decontouring} or selective average filter \cite{song2016hardware}, or even directly adding noise to alleviate the artifacts \cite{daly2003bit}. Learning-based methods \cite{hou2017image, liu2018learning, zhao2019deep, punnappurath2020little} have been proposed recently and they usually focus on restoration from lower bit-depth input to the 8-bit image. In this work, we aiming at recovering a 16-bit HDR image from an 8-bit LDR image.

\section{Methodology}
\subsection{Observations}\label{sec:Observations}
\begin{figure}[htp]
    \begin{center}
    \includegraphics[width=1\linewidth]{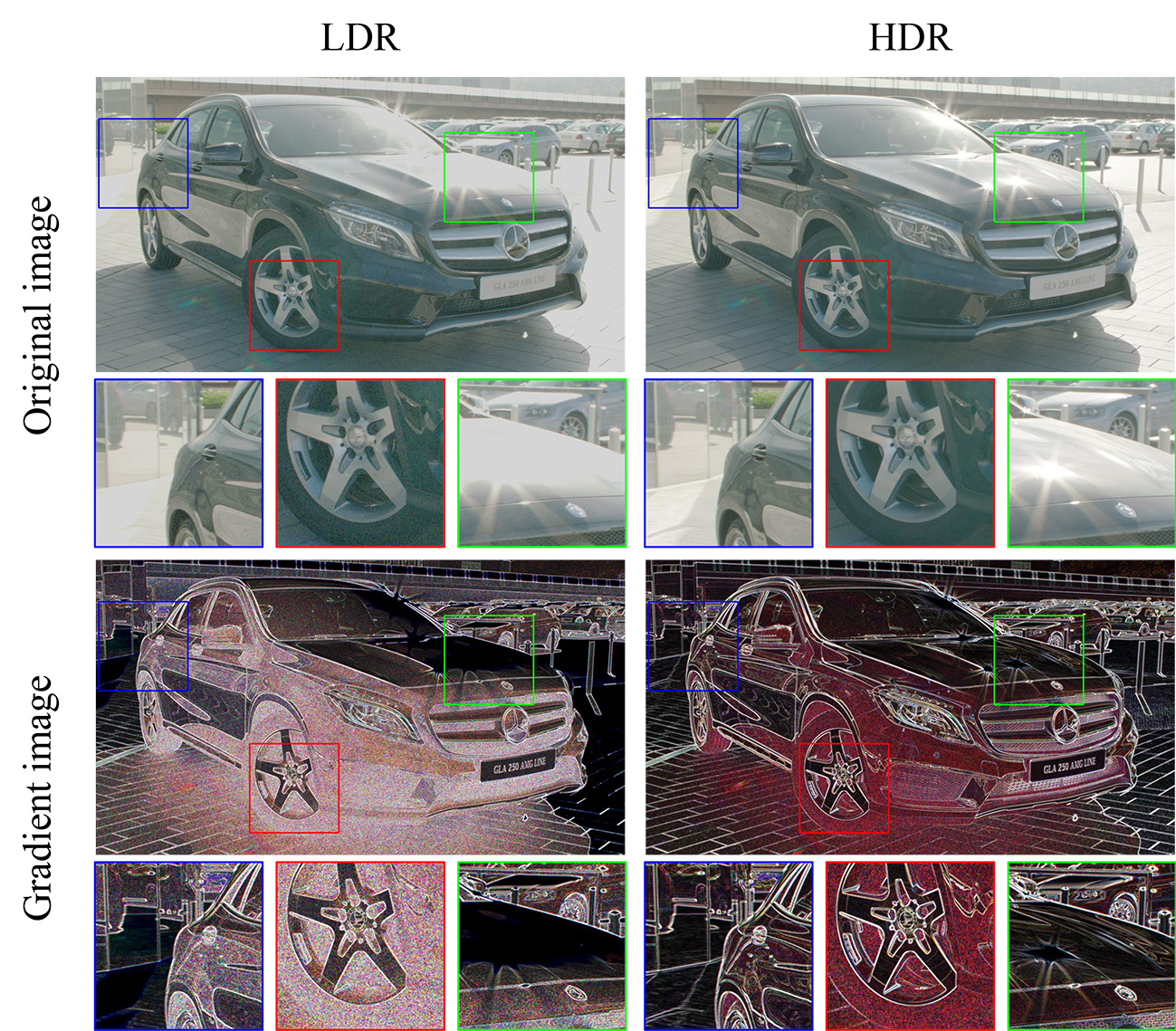}
    \end{center}
    \vspace{-7pt}
    \caption{Gradient maps calculated by Scharr operator \cite{scharr2004optimal} of the LDR and HDR image. Note that the HDR image is not noise-free. It can be observed that gradients of LDR and the corresponding HDR image are obviously different both in over-exposed regions and well-exposed regions.}
    \vspace{-15pt}
    \label{Fig: observation}
\end{figure}

The problem of image HDR reconstruction is often accompanied with denoising and dequantization. To illustrate this point, we visualize the gradient map of an LDR image and the corresponding HDR image by Scharr operator \cite{scharr2004optimal} as shown in Figure \ref{Fig: observation}. Compared with the HDR image, gradients are less visible in highlight areas of the LDR image, due to the dynamic range compression and quantization. In the well-exposed areas, gradients of noises are clear in the LDR and HDR images, indicating that noise exists in both images. Nevertheless, patterns of noise are markedly different between LDR and HDR images due to different noise levels. In addition, unlike Gaussian white noise that is uniformly distributed throughout the whole image, distribution of noises in these images are not uniform. Therefore, the pattern difference does not only exist between the highlight and non-highlight areas, but also in different positions of well-exposed regions. This inspires us to design a spatial-variant modulation module for the network.

\subsection{Network Structure}
\begin{figure*}[!t]
    \begin{center}
    \includegraphics[width=0.95\linewidth]{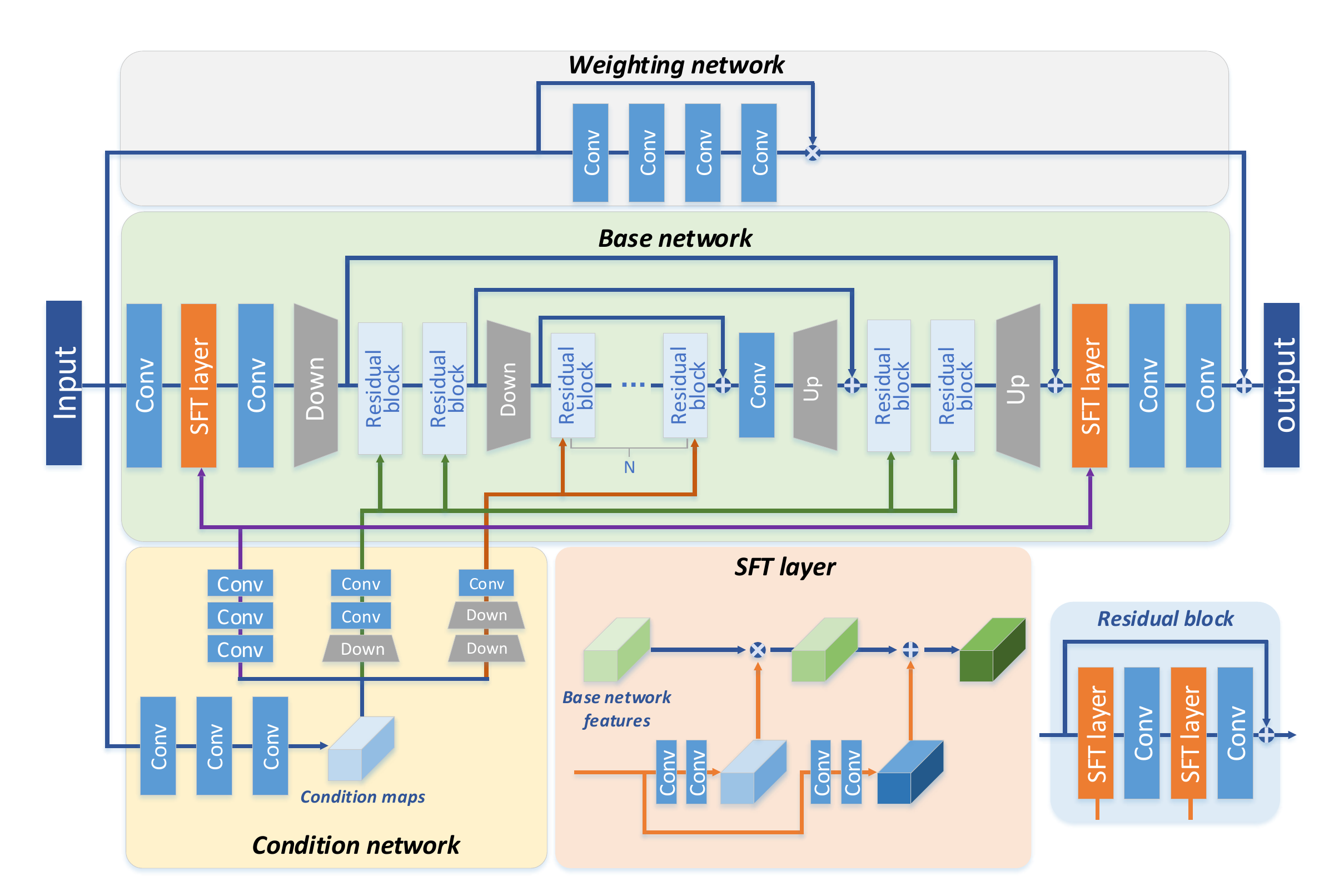}
    \end{center}
    \vspace{-15pt}
    \caption{Network structure of our HDRUNet with a base network, a condition network and a weighting network. The three modules all take the LDR image as input. Particularly, the condition network predicts condition maps that afterwards utilized to modulate the intermediate features in the base network.}
    \label{Fig: Network Structure}
    \vspace{-10pt}
\end{figure*}

Based on the aforementioned observation, we design a UNet-like network with spatial modulation for the single image HDR reconstruction. The overall architecture of the proposed method is depicted in Figure \ref{Fig: Network Structure}, which consists of three main components -- a base network, a condition network and a weighting network.

\textbf{Base Network.}
The base network utilizes a UNet-like structure, which takes the 8-bit noisy LDR image as input and reconstructs the 16-bit HDR image. The predicted HDR images are supposed to contain more details in under-/over-exposed areas with little noise. Many image reconstruction algorithms \cite{eilertsen2017hdr, liu2020single} have proven the effectiveness of UNet-like structure, which can make full use of the hierarchical multi-scale information from low-level features to high-level features. We adopt similar concept for this task. The encoder is devised to map the LDR image to high-dimensional representations, and the decoder is trained to reconstruct the HDR image from the encoded representations. To achieve better reconstruction performance, skip connections are added between the encoder and decoder. In the task of HDR reconstruction, the encoder and decoder work in 8-bit and 16-bit, respectively. To ease the training procedure and maximize the information flow, several residual blocks are utilized in the base network.

\textbf{Condition Network.} The key to reconstruct HDR images is to recover the missing details in under-/over-exposed regions of the input LDR image. Different areas in one image have different exposures and brightness. Further, various images also have different holistic brightness and contrast information. Hence, it is necessary to deal with input images with location-specific and image-specific operations. Besides, non-uniformly distributed noise also requires the network to process various patterns well. However, conventional convolutional neural networks are spatially variant, where the same filter weights are applied across all images and local regions. Thus, inspired by \cite{wang2018recovering, liu2021very}, we introduce a condition network with spatial feature transform (SFT) \cite{wang2018recovering} to provide spatially variant manipulations. Specifically, the condition network accepts the input LDR image and predicts the corresponding conditional maps that are afterwards used to modulate the intermediate features in the base network. The structure of the condition network and the mechanism of SFT layer are portrayed in Figure \ref{Fig: Network Structure}.
\begin{equation}
SFT(x) = \alpha \odot x + \beta,
\end{equation}
where $\odot$ denotes the element-wise multiplication. $x \in\mathbb{R}^{C \times H \times W} $ is the intermediate features to be modulated. $\alpha \in\mathbb{R}^{C \times H \times W}$ and $\beta \in\mathbb{R}^{C \times H \times W}$ are two modulation coefficient maps predicted by the condition network. By leveraging such modulation strategy, our method can achieve location and image specific manipulation according to different inputs. Experiments have demonstrated the effectiveness of such feature modulation for HDR reconstruction with denoising and dequantization.

\textbf{Weighting Network.} The biggest challenge of HDR reconstruction is to restore fine details in under-/over-exposed regions, while most of the well-exposed contents can be of less contribution to the learning procedure. To this end, we propose a weighting estimation network to forecast a soft weighting map $W$ on the well-exposed regions to be retained. Thereupon, the whole network will pay more attention to reconstruct the details of over-exposed areas.
\begin{equation}
\hat{Y} = W \odot I + \mathcal{G}(I),
\end{equation}
where $I$ is the input LDR image, $\hat{Y}$ is the final reconstructed HDR image, and $\mathcal{G}(I)$ is the output of the base network.

\subsection{Loss Function}\label{sec:loss_function}
In real-world image HDR reconstruction, it is necessary to consider not only the restoration of the dynamic range, but also the reduction of noise and quantization artifacts. However, loss functions that are commonly used in previous works of image restoration, such as $L_1$ and $L_2$ loss, are not applicable to simultaneously deal with these aforementioned problems. A loss function formulated directly on HDR values will make the network focus on high luminance values and underestimate the impact in lower luminance values, resulting in worse quantitative performance and visual quality. The experimental results can be found in Section \ref{sec:ablation_study}. Therefore, we propose a specially designed $Tanh \textunderscore L_1$ loss for the task, which is formulated as:
\begin{equation}
	Tanh \textunderscore L_1 (\hat{Y}, H) = \left| Tanh (\hat{Y})-Tanh (H) \right|,
\end{equation} where $\hat{Y}$ and $H$ represent the predicted HDR image and the corresponding ground truth image, respectively.

\section{Experiments}
\subsection{Experimental Setup}\label{sec:experimental_setup}
\textbf{Dataset.} Previous studies \cite{endo2017deep, eilertsen2017hdr, liu2020single, kim2019deep} have adopted different datasets on the task of image HDR reconstruction for training and evaluation. In this paper, we use the dataset proposed by NITRE 2021 HDR Challenge \cite{perez2021ntire}. As depicted in this challenge, the dataset is a subset of images selected from the HdM HDR dataset \cite{froehlich2014creating}, where the HDR images are captured by two Alexa Arri cameras with a mirror rig and the corresponding LDR images are generated by applying a degradation model (e.g., exposure gain, noise addition and quantization, clipping). In this dataset, there are 1494 LDR/HDR pairs for training, 60 images for validation and 201 images for testing. Note that the LDR/HDR pairs are aligned both in time axis and exposure level and stored after gamma correction (i.e., they are non-linear images). Since the ground truths of the validation and testing set are not available, we conduct the experiments only based on the training set. The training set is composed of 1494 consecutive frames in 26 long takes. We randomly select 3 frames in every long take, a total of 78 frames, as the verification set, and the rest 1416 frames are used for training. 

\textbf{Evaluation Metrics.} In the challenge, standard PSNR directly computed in the output images (normalized to the peak value of the ground-truth HDR image) and PSNR computed in the $\mu$-law tone-mapped images (normalized to the 99 percentile of the ground-truth image and bounded by a Tanh function to avoid excessive brightness compression) are used as the evaluation metrics. We represent these two metrics as PSNR-L and PSNR-$\mu$, respectively. It can be seen that PSNR-L and PSNR-$\mu$ have different tendencies for evaluating image quality. For s-PNSR, the accuracy of highlight values is the most important influential factor. However, these values are often severely compressed by tone mapping for visualization. While PSNR-$\mu$ directly measures the tone-mapped values that can directly reflect the visual similarity of the result and the ground truth. Therefore, the main measure in quantitative comparisons is PSNR-$\mu$ both in the challenge and in this paper.

\textbf{Implementation Details.} In the following experiments, the number of residual blocks $N$ is set to 8. Convolution filters with stride of 2 are used for down-sampling and pixel shuffle \cite{shi2016real} is utilized for up-sampling. Before training, we pre-process the data by cropping images into $480\times 480$ with step of 240. During training, the mini-batch size is set to 16 and the number of training iterations is set to $1\times 10^{6}$. Adam \cite{kingma2014adam} optimizer and Kaiming-initialization \cite{he2015delving} are adopted for training. The initial learning rate is set to $2\times 10^{-4}$ and decayed by a factor of 2 after every $2\times 10^{5}$ iterations. All models are built on the PyTorch framework and trained with NVIDIA 2080Ti GPU. When the patch size of input is set to $256 \times 256$, the total training time is about 5 days.

\subsection{Ablation Study}\label{sec:ablation_study}
In this section, we conduct ablation study to further investigate the different settings, including the training patch size, loss functions, key modules and modulation strategies.

\textbf{Training Patch Size.}
In practice, we find that the training patch size has an important influence on this task. In general,  small patch size (e.g., $32 \times 32$ or $64 \times 64$) is usually adopted during training in super-resolution networks \cite{dong2014learning, wang2019edvr}. However, HDR reconstruction is more than a simple local process. It involves more global and holistic manipulations, since different regions in LDR image require different treatments. Besides, due to severe information loss in over-exposed regions, we believe that restoration of the details needs a large receptive field in these areas. As shown in Table \ref{tab:patch_size}, with the increase of patch size, the quantitative performance is gradually improved. To consider both performance and computational cost, we select $256 \times 256$ as the recommended patch size.

\renewcommand\arraystretch{1.3}
\begin{table}[htbp]
	\begin{center}
		\begin{tabular}{ccc}
			\toprule
			Patch size & PSNR-L (dB) & PSNR-$\mu$ (dB) \\ \hline
			48 & 39.82 & 33.43 \\ 
			96 & 40.60 & 33.78 \\ 
			160 & 41.13 & 33.94 \\ 
			256 & 41.61 & 34.02 \\
			\bottomrule
		\end{tabular}
		
	\end{center}
	%\vspace{-7pt}
	\caption{Influence of training patch size.}
	\label{tab:patch_size}
	\vspace{-5pt}
\end{table}

\textbf{Loss Function.}
In Section \ref{sec:loss_function}, we introduce a $Tanh \textunderscore L_1$ loss for HDR reconstruction with denoising and dequantization. To accelerate the training process, we fix the patch size to $160 \times 160$. To validate the effectiveness of our proposed loss function, we conduct experiments with various loss functions and make quantitative and qualitative comparisons. The quantitative results are shown in Table \ref{tab:loss_function}, from which we can draw the following observations: 1) Compared with $L_2$ loss, $L_1$ loss can obtain better quantitative performance with higher PSNR-L and PSNR-$\mu$ values. 2) By introducing $Tanh$ operation, the PSNR-$\mu$ can be further improved at the cost of PSNR-L. To be specific, using $Tanh \textunderscore L_1$ loss improves PSNR-$\mu$ by 0.35 dB. This is because when $L_1$ or $L_2$ loss function is used directly, the training loss of the high brightness value has larger weight. In this case, the network mainly focuses on the highlight areas, leading to higher PSNR-L. However, as depicted in Section \ref{sec:experimental_setup}, PSNR-$\mu$ can better reflect the visual similarity of the output with the ground truth. Since the PSNR-$\mu$ is also the main reference evaluation metric in the challenge, we adopt $Tanh \textunderscore L_1$ as the loss function. 

\begin{table}[htbp]
	\begin{center}
		\begin{tabular}{ccc}
	    \toprule
			Loss & PSNR-L (dB) & PSNR-$\mu$ (dB) \\ \hline
			$L_2$ & 43.91 & 31.80 \\ %\hline
			$L_1$ & 44.10 & 33.59 \\ %\hline
			$Tanh{-}L_2$ & 40.02 & 33.92 \\ %\hline
			$Tanh \textunderscore L_1$ & 41.13 & 33.94 \\ 
		\bottomrule
		\end{tabular}
	\end{center}

	\caption{Quantitative comparison of different loss functions.}
	\label{tab:loss_function}
	\vspace{-7pt}
\end{table}

\begin{figure}[htp]
    \begin{center}
    \includegraphics[width=1\linewidth]{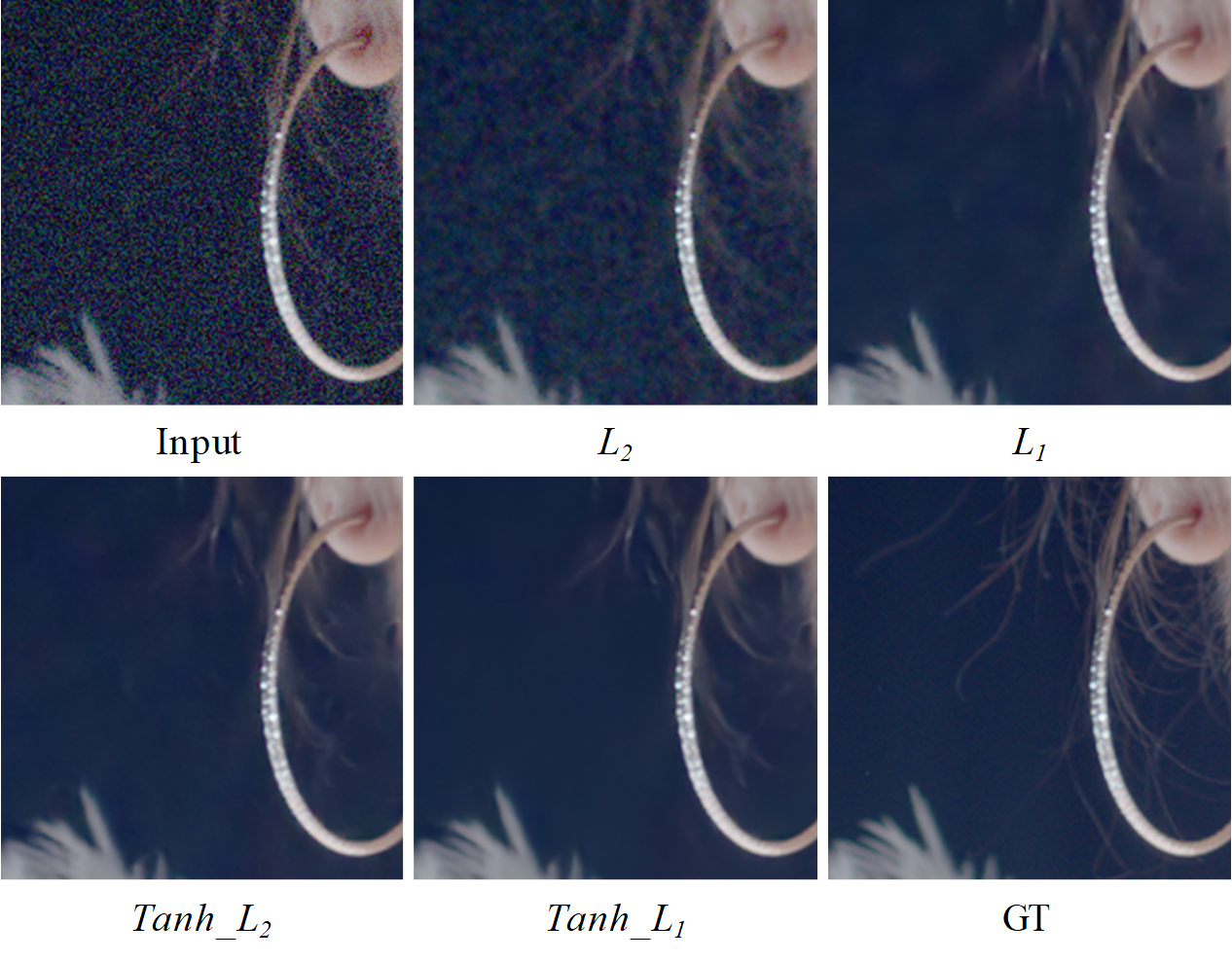}
    \end{center}
    \caption{Visual comparison of different loss functions. It can be obviously observed that $L_1$ or $L_2$ loss perform badly for denoising, and our $Tanh \textunderscore L_1$ loss achieves the best visual quality.}
    \label{Fig: loss}
\end{figure}

Moreover, the loss function also has a significant impact on the visual results. The visual comparison of these loss functions are shown in Figure \ref{Fig: loss}. We can see that results generated by using $L_1$ or $L_2$ loss function perform badly for denoising in well-exposed regions. In contrast, $Tanh \textunderscore L_1$ loss achieves the best visual quality.

\textbf{Effectiveness of Key Modules.}
In this section, we demonstrate the effectiveness of each proposed component. The experimental results are shown in Table \ref{tab:component}. Note that we set patch size of $160 \times 160$ for fast training. If we only adopt a sole UNet-like base network, the PSNR-L and PSNR-$\mu$ are 40.77 dB and 33.85 dB, respectively. By adopting the weighting network branch, the performance is slightly improved. If we combine the base network and the condition network together, the PSNR-L and PSNR-$\mu$ are improved by 0.27 dB and 0.06 dB. With all three key modules equipped, our full model can further achieve higher quantitative results with PSNR-L of 41.13 dB and PSNR-$\mu$ of 33.94 dB. The results clearly validate the effectiveness of the proposed key modules.

\begin{table}[htbp]
	\begin{center}
		\begin{tabular}{ccccc}
			\toprule
			Network Structure & \multicolumn{4}{c}{Base Network} \\ \hline
			Condition Network & \text{\sffamily X} & \text{\sffamily X} & \checkmark & \checkmark \\ 
			Weighting Network & \text{\sffamily X} & \checkmark & \text{\sffamily X} & \checkmark \\ 
			\midrule
			PSNR-L (dB) & 40.77 & 40.85 & 41.04 & 41.13\\ 
		    PSNR-$\mu$ (dB) & 33.85 & 33.90 & 33.91 & 33.94\\ 
			\bottomrule
		\end{tabular}
		
	\end{center}
	
	\caption{Effectiveness of each proposed component.}
	\label{tab:component}
	%\vspace{-10pt}
\end{table}

\textbf{Exploration on Modulation Strategy.}
Feature modulation has proven to be an effective way to tackle image-specific and location-specific tasks, such as photo retouching \cite{he2020conditional}, image restoration \cite{he2019modulating, he2019interactive}, image super-resolution \cite{wang2018recovering}, as well as HDR reconstruction \cite{kim2019deep}. In this paper, we adopt SFT to provide spatially variant manipulations. We also compare other feature modulation vairants. In our condition network, the size of the predicted condition maps is $C \times H \times W$, thus, every unit of the feature maps in the based network will be modulated. The condition maps can also be of size $1 \times H \times W$, in which case the modulation parameters are spatial-variant but shared across channels. In contrast, the modulation in CResMD \cite{he2019interactive} is global channel-wise without considering spatial information.

\begin{table}[htbp]
	\begin{center}
		\begin{tabular}{ccc}
			\toprule
			Modulation strategy & PSNR-L (dB) & PSNR-$\mu$ (dB) \\ \hline
			None & 40.77 & 33.85 \\ 
			CResMD ($C \times 1 \times 1$) & 39.84 & 33.65 \\ 
			SFT ($1 \times H \times W$) & 40.65 & 33.82 \\ 
			SFT ($C \times H \times W$) & 41.04 & 33.91 \\ 
			\bottomrule
		\end{tabular}
	\end{center}
	\caption{Comparison of different modulation strategies.}
	\label{tab:modulation}
\end{table}

\begin{figure*}[htbp]
    \begin{center}
    \includegraphics[width=0.91\linewidth]{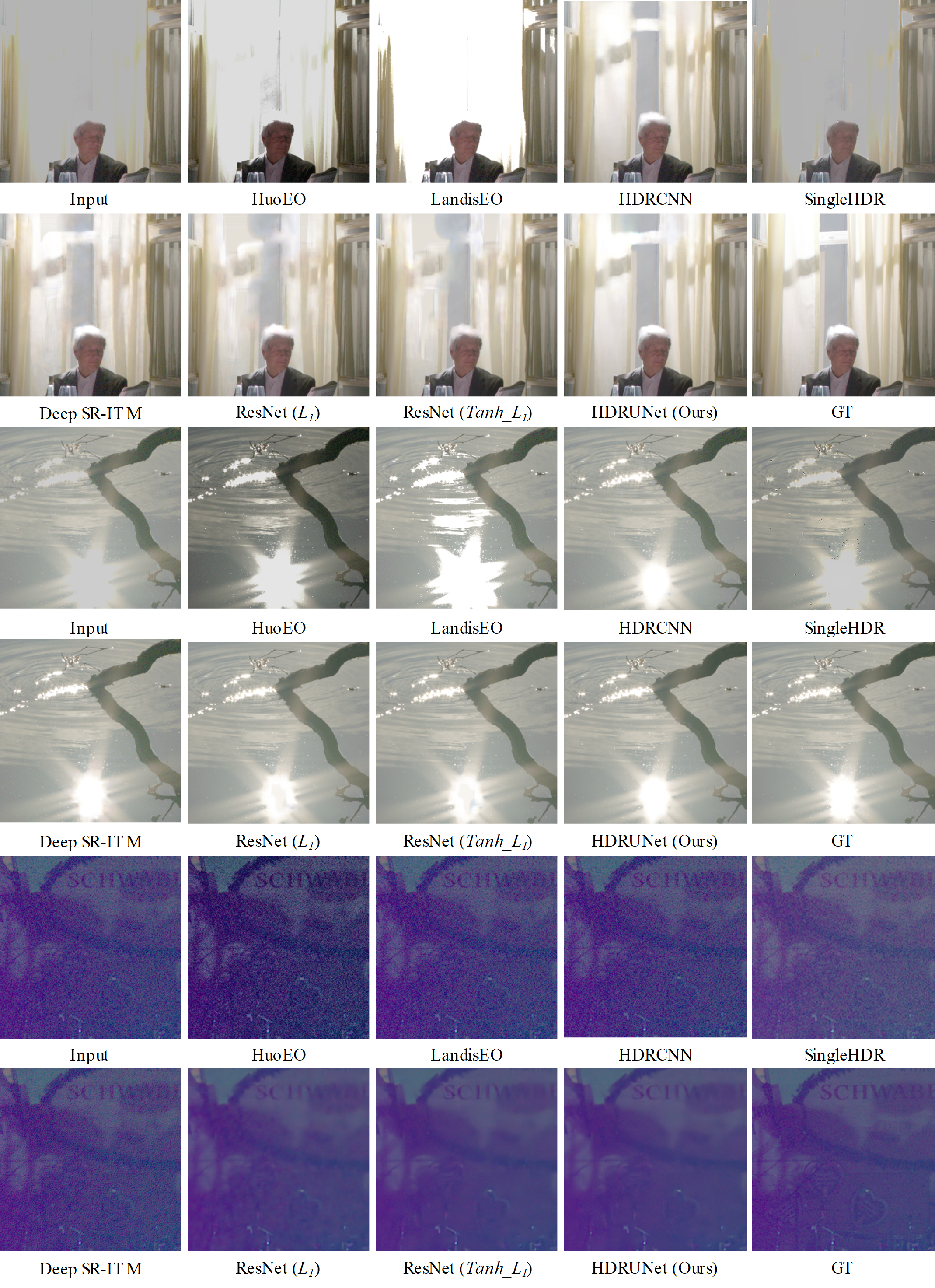}
    \end{center}
    \caption{Qualitative comparison with other methods.}
    \label{Fig: main_exp}
\end{figure*}

The comparison results of these modulation strategies are listed in Table \ref{tab:modulation}. Note that, to directly illustrate the differences among various modulation methods, we eliminate the weighting network in the experiments. From the results, it can be observed that global channel-wise modulation has little effect on HDR reconstruction, since it cannot provide any spatially variant manipulation. By introducing SFT, the performance is greatly improved, which validate our comments that different areas in LDR image should be handled differently. Moreover, spatial modulation with $C \times H \times W$ is superior to that with $1 \times H \times W$, since it cannot only involve spatial-wise but also channel-wise manipulations.

\subsection{Comparison with State-of-the-art Methods}
We compare our HDRUNet with several state-of-the-art methods on image HDR reconstruction, including LandisEO \cite{landis2002production}, HuoEO \cite{huo2013dodging}, HDRCNN \cite{eilertsen2017hdr}, SingleHDR \cite{liu2020single}, Deep SR-ITM \cite{kim2019deep} and a ResNet-style \cite{he2016deep} network. However, most of these methods utilize different datasets that contain many specific operations for the data to be processed. We slightly modify these algorithms or add post-processing for this dataset. For LandisEO and HuoEO, we use the implementations in HDR Toolbox \cite{Banterle:2017} and set the gamma as 2.24. Besides, we implement gamma correction on the results because these are linear HDR values. For HDRCNN, we retrain them on the same dataset as our method. We use a convolution filter with stride of 2 for down-sampling to match the size of input and output for Deep SR-ITM. Since SingleHDR is only suitable for restoring the linear HDR value in luminance domain, we directly test the pretrained model on our dataset and implement post-process as LandisEO and HuoEO. We also train a ResNet-style model that is commonly used in image restoration and utilize both $L_1$ loss and the proposed $Tanh \textunderscore L_1$ loss.

\begin{table}[htbp]
	\begin{center}
		\begin{tabular}{ccc}
			\toprule
			Method & PSNR-L (dB) & PSNR-$\mu$ (dB) \\ \hline
			LandisEO & 17.88 & 23.30 \\ 
			HuoEO & 32.40 & 17.35 \\ 
			SingleHDR & 32.32 & 19.54 \\ 
			HDRCNN & 39.47 & 26.05 \\ 
			Deep SR-ITM & 43.29 & 26.25 \\ 
			ResNet ($L_1$) & 41.92 & 33.24 \\ 
			ResNet ($Tanh \textunderscore L_1$) & 39.82 & 33.67 \\ 
			HDRUNet (Ours) & 41.61 & 34.02 \\ 
			\bottomrule
		\end{tabular}
		
	\end{center}
	\caption{Quantitative comparison with other methods.}
	\label{tab:Quantitative Comparison}
\end{table}

\textbf{Quantitative Comparison.}
We provide the quantitative results in Table \ref{tab:Quantitative Comparison}. As described in Section \ref{sec:experimental_setup}, PSNR-L and PSNR-$\mu$ have different tendencies for evaluating image quality. PSNR-L is used to measure the accuracy of the high luminance values, while PSNR-$\mu$ reflects the visual similarity between predicted HDR image and the ground truth. For LandisEO, HuoEO and SingleHDR, these methods predict linear HDR values. Although we perform gamma correction on the results, it is still very difficult to align the exposure completely. Thus the results generated by these methods perform badly in such reference-based metrics. HDRCNN and Deep SR-ITM learn direct mapping from LDR to HDR, while HDRCNN only processes values in over-exposed regions. Deep SR-ITM uses a big network with $L_2$ loss function, which brings higher PSNR-L and lower PSNR-$\mu$. It can be seen that our method achieves the best quantitative performance in PSNR-$\mu$ and far above average performance in PSNR-L.

\textbf{Qualitative Comparison.}
We provide the qualitative comparison in Figure \ref{Fig: main_exp}. Our method can not only restore fine details in highlight regions but also greatly reduce noise in lower luminance area. On the contrary, although the other methods improve the brightness of the highlight areas, they hardly recover details in these areas and some of them introduce additional artifacts. LandisEO uses a global operator to increase the brightness in over-exposed regions but can not generate details. HuoEO and ResNet generate some details but introduce additional artifacts at the same time. Additionally, these methods can not preform denoising and dequantization well. Noise can be clearly observed in the well-exposed regions. In comparison of these approaches, results of our method achieve the best visual quality.  

\subsection{Results of NTIRE2021 HDR Challenge}
We participated in the NTIRE2021 HDR Challenge \cite{perez2021ntire} and won the second place in the single frame track. The results are shown in Table \ref{tab:competition results}. Without using ensemble approaches, our method obtains similar PSNR-$\mu$ score as the first place, only about $0.07$ dB apart. Besides, the running speed of ours is more than 116 times that of the first place.

\begin{table}[htbp]
	\begin{center}
% 	\scalebox{0.78}{
    \setlength\tabcolsep{1pt}{
		\begin{tabular}{ccccc}
			\toprule
			Team & PSNR-$\mu$ & PSNR-L & Runtime (s) & Ensemble \\ \hline
			NOAHTCV & 34.804 & 32.867 & 61.52 & \checkmark \\
			XPixel (ours) & 34.736 & 32.285 & 0.53 & - \\
			BOE-IOT-AIBD & 34.414 & 33.490 & 5.00 & - \\
			CET CVLab & 33.874 & 32.06 & 0.20 & \checkmark \\
			CVRG & 32.778 & 31.021 & 1.10 & - \\
			\bottomrule
		\end{tabular}
	}
	\end{center}
	\caption{Results of the top5 methods in the challenge.}
	\vspace{-10pt}
	\label{tab:competition results}
\end{table}

\section{Conclusion}
In this paper, we propose a spatially dynamic encoder-decoder network, HDRUNet, with a novel $Tanh \textunderscore L_1$ loss function to solve the single image HDR reconstruction problem. Our method won the second place in the single frame track of NTIRE2021 HDR Challenge. Particularly, the proposed network contains three modules which are a base network, a condition network and a weighting network. The base network can exploit multi-scale information to reconstruct HDR image. The condition network makes use of SFT layer to perform spatial-variant modulation for various patterns. The weighting network can retain useful information of the input LDR image for helping learning. Moreover, we introduce a $Tanh \textunderscore L_1$ loss to balance the weight of learning for high luminance values and the other values. Using this function greatly facilitates learning for joint HDR reconstruction with denoising and dequantization. Overall, our methods outperforms state-of-the-art methods in quantitative and qualitative comparisons.

\textbf{Acknowledgement.} This work was supported in part by the Shanghai Committee of Science and Technology, China (Grant No. 20DZ1100800), in part by the National Natural Science Foundation of China under Grant (61906184), Science and Technology Service Network Initiative of Chinese Academy of Sciences (KFJSTSQYZX092), Shenzhen Institute of Artificial Intelligence and Robotics for Society.

\clearpage
{\small
\bibliographystyle{ieee_fullname}
\bibliography{egbib}
}

\end{document}